\documentclass[journal, letterpaper, twocolumn]{IEEEtran}

\usepackage[T1]{fontenc}
\usepackage[latin9]{inputenc}
\usepackage{algorithm}
\usepackage{amsmath}
\usepackage{amssymb}
\usepackage{graphicx}
\usepackage{setspace}
\usepackage{bm}
\usepackage{algpseudocode}
\usepackage{array,multirow,makecell}\setcellgapes{1pt}
\usepackage[table]{xcolor}
\usepackage{setspace}

\newcolumntype{R}[1]{>{\raggedleft\arraybackslash }b{#1}}
\newcolumntype{L}[1]{>{\raggedright\arraybackslash }b{#1}}
\newcolumntype{C}[1]{>{\centering\arraybackslash }b{#1}}
\makegapedcells

\renewcommand\IEEEkeywordsname{Index Terms}

\begin{document}
\pagestyle{empty}

\title{\huge Joint Optimal Number of RF chains and Power Allocation for Downlink Massive MIMO Systems}
\author{\IEEEauthorblockN{Rami Hamdi $^{1,2}$ and Wessam Ajib $^{2}$} \\
\IEEEauthorblockA{$^{1}$ Department of Electrical Engineering, École de Technologie Supérieure  \\ rami.hamdi.1@ens.etsmtl.ca \\
$^{2}$ Department of Computer Science, Université du Québec à Montréal \\
ajib.wessam@uqam.ca }
}
\maketitle
\thispagestyle{empty}

\begin{abstract}
  This paper investigates the downlink of massive multiple-input multiple-output (MIMO) systems that include a single cell Base Station (BS) equipped with large number of antennas serving multiple users. As the number of RF chains is getting large, the system model considered in this paper assumes a non negligible circuit power consumption. Hence, the aim of this work is to find the optimal balance between the power consumed by the RF chains and the transmitted power. First, assuming an equal power allocation among users, the optimal number of RF chains to be activated is analytically found. Then, for a given number of RF chains we derive analytically the optimal power allocation among users. Based on these analysis, we propose an iterative algorithm that computes jointly the optimal number of RF chains and the optimal power allocation vector. Simulations validate the analytical results and show the high performance provided by the proposed algorithm.
\end{abstract}

\section{Introduction}
The recent huge increase of utilization of wireless transmission technologies creates the need to innovate and find new technologies that offer higher performance. In the fifth generation (5G) of cellular networks, one of the main objectives is to increase by several orders of magnitude the rate offered by the current cellular systems. Massive multiple-input multiple-output (MIMO) system is a new paradigm based on the multiplexing of few hundred of antennas serving at the same time-frequency few tens of users~\cite{five,five2}. This paradigm is largely accepted as a candidate for the design of future 5G cellular networks since it provides several gains, such as high throughput~\cite{over1}. These gains are obtained while using a limited transmission power as well as maintaining a reduced complexity of the transmission and receiving systems~\cite{over2}.

Massive MIMO gains can not be efficiently exploited without adequate allocation of resources. Some previous works investigated how to optimize the resource allocation for massive MIMO systems under different network architectures and assumptions. In~\cite{res1}, the authors show that massive MIMO technology provides a good trade-off between energy efficiency and spectral efficiency for both prefect and imperfect Channel State Information (CSI). This good performance can be achieved under simple linear beamformers such as Zero Forcing (ZF) and conjugate beamforming. In~\cite{res2,res3}, the optimal beam forming vector that maximizes the sum-rate under the sum-power constraint is found. Antenna selection strategies are investigated for massive MIMO systems in some works. In~\cite{as1}, the authors propose a joint antenna selection and power allocation technique that maximizes the sum-rate. A distributed massive MIMO systems with limited backhaul capacity is investigated in~\cite{as2}. The authors propose a joint antenna selection and user scheduling strategy. While in~\cite{as3}, an efficient algorithm is proposed for downlink massive MIMO systems when assuming a limited number of RF chains. This algorithm  allows to jointly select antennas and schedule users with low computational complexity.

The specificity of massive MIMO systems is the use of large number of antennas. As a consequence, there is a significant amount of power needed for the operation of RF chains. The circuit power consumption is often neglected. Anyhow, It has been considered in few previous works~\cite{cir2,cir1}. Based on~\cite{cir1}, it is not true that more RF chains implies always higher rate. In this paper, we investigate the optimal number of RF chains that maximizes the sum-rate in a massive MIMO system. We assume simple random antenna selection in this paper. More performant antenna selection schemes will be investigated in future works. First, we express analytically the optimal number of RF chains in the case of equal power allocation among users. Next, we find the optimal power allocation vector for fixed number of RF chains. Moreover, based on such results, we propose an algorithm that jointly computes the optimal number of RF chains and the optimal power allocation among users. The proposed algorithm allows to find the optimal balance between the amount of power consumed by RF chains and the transmitted power. Analytical results are validated by simulations.

In this paper, all power variables are assumed to be unitless since they are normalized by the average noise power. $\mathbf{E}\{.\}$ denotes the mathematical expectation, $\lfloor.\rfloor$ denotes the floor function, $\lceil.\rceil$ denotes the ceiling function, $(x)^+$ denotes $\text{max}(0,x)$, $(.)^H$ represents the Hermitian matrix, $\mid.\mid$ represents the Euclidean norm of a vector and $\parallel.\parallel_{F}$ denotes the Frobenius norm of a matrix.

The paper is organized as follows. In Section II, the system model is presented and the problem is formulated. In Section III, the optimal number of RF chains is analytically calculated when transmit power is assumed equally allocated among users. Then, we describe the proposed algorithm that optimizes joinly the number of RF chains and power allocation among users in Section IV. Numerical and simulation results are shown and discussed in Section V. Finally, we conclude the main results in Section VI.

\section{System Model and Problem Formulation}
This paper considers the downlink of a single cell massive MIMO systems. The base station (BS) is equipped with a large number of antennas $N$ serving $K$ single-antenna users with $N \gg K$. Considering only small scale fading and under favorable propagation condition, the complex channel gain matrix $\bm{H}=[\mathbf{h_1}, \mathbf{h_2}, ... ,\mathbf{h_K}]$, where $\mathbf{h_k} \in \mathbb{C}^{1\times N}$ is the $k^{th}$ slow fading channel vector for user $k$, is assumed to be quasi-static independent and identically distributed (i. i. d.) complex Gaussian random variables with zero mean and unit variance. The noise is assumed to be additive Gaussian random variable with zero mean and unit variance. We consider that the BS knows perfectly the Channel State Information (CSI). The beam forming matrix is $\bm{W}=[\mathbf{w_1} \mathbf{w_2} ... \mathbf{w_K}]$, where $\mathbf{w_k} \in \mathbb{C}^{N\times1}$ is the $k^{th}$ beam forming vector for user $k$.
The vector $\mathbf{p}=[p_1 p_2 ... p_K]$ is assumed to present the values of power allocated to users. The received downlink vector of signals can be expressed as

\begin{equation}
 \label{eq:1}
   \mathbf{y}= \bm{H} \mathbf{x}+\mathbf{n},
\end{equation}
where the transmitted signal is given by

\begin{equation}
 \label{eq:2}
   \mathbf{x}=\bm{W} diag(\sqrt{\mathbf{p}}) \mathbf{s},
\end{equation}
where $\mathbf{s}$ is the data symbol vector with unit energy. The signal received by user $k$ can be written as

\begin{equation}
 \label{eq:3}
   y_k= \sqrt{p_k} \mathbf{h_k} \mathbf{w_k} s_k+ \sum_{i=1,i\neq k}^{K} \sqrt{p_i} \mathbf{h_k} \mathbf{w_i} s_i+n_k
\end{equation}

Without loss of generality, we consider conjugate beamfoming pre-coder because of its great performance and low complexity. Moreover, it is shown that conjugate beamforming achieves high spectral efficiency~\cite{mrt}. The Zero Forcing (ZF) and Regularized Zero Forcing (RZF) can be considered in future works. Hence, the conjugate beamforming matrix is given as $\bm{W}=\bm{H}^{H} / \parallel\bm{H}^H\parallel_{F}$. The received $SINR$ for user $k$ is expressed as

\begin{equation}
 \label{eq:4}
   {SINR}_k=\frac{\frac{p_k}{\eta}\mid\mathbf{h_k}\mathbf{h_k}^H\mid^2}{\sum_{i=1,i\neq k}^{K}\frac{p_i}{\eta}\mid\mathbf{h_k}\mathbf{h_i}^H\mid^2 +1}
\end{equation}
where $\eta=\parallel\bm{H}^{H}\parallel_{F}^2$.

The rate for user $k$ is given as

\begin{equation}
 \label{eq:5}
   R_k=\log_2(1+{SINR}_k)
\end{equation}

Now, we introduce the circuit power consumption model as in~\cite{cir2,cir1}. Let $p_c$ denotes the amount of power consumed at each RF chain (Digital to Analog Converter (DAC), mixer, frequency synthesizer, filter). The value of $p_c$ is assumed independent from the output transmitted power $p_{out}$~\cite{cir1}. Let $p_{max}$ denotes the maximum power available at the BS. So the circuit power constraint can be expressed as

\begin{equation}
\label{eq:7}
p_{out}+\sum_{n=1}^{N}\alpha_n . p_c \leq p_{max}
\end{equation}
where $\alpha_n$ is an antenna coefficient that is set to 1 if antenna $n$ is selected and to 0 otherwise. Hence, $p_{out}$, the output transmitted power, is given by

\begin{equation}
\label{eq:8}
p_{out}=\sum_{k=1}^{K}p_k
\end{equation}

The circuit power consumption causes that the achieved rate is not necessarily maximized when all the RF chains available at the transmitter are activated~\cite{cir1}. Hence, the optimal number of RF chains that maximizes the sum-rate should be derived. Moreover, the main problem is to find the optimal balance between the portion of power consumed at the RF chains and the portion consumed for transmitting. Hence, the optimal number of RF chains and the optimal power distribution among users have to be jointly derived to achieve the maximum average sum-rate. The sum-rate is averaged over $\bm{H}$ realizations. The problem can be formulated as

\begin{equation}
\label{eq:9}
\begin{aligned}
& \underset{\mathbf{p},\alpha_n:n=1..N}{\text{maximize}}
& & \overline{R}= \mathbf{E}\{\sum_{k=1}^{K} R_k \} \\
& \text{subject to}
& & \sum_{k=1}^{K}p_k+\sum_{n=1}^{N}\alpha_n . p_c \leq p_{max}, \\
&&& \sum_{n=1}^{N}\alpha_n \geq K, \\
&&& \alpha_n \in \{0,1\},n=1..N.
\end{aligned}
\end{equation}

The objective function is non-convex due to the multi user interference. Hence, the formulated problem is non-convex and water filling is not the optimal strategy for power allocation among users~\cite{wf}. In the next section, the problem given in (\ref{eq:9}) is simplified. The sum-rate is approximated when the transmit power is equally allocated among users and the optimal number of RF chains is given.

\section{Optimal Number of RF Chains when Equal Power Allocation}

In this section, equal power allocation among users is assumed and hence $p_k=p_{out}/K,k=1:K$. The aim is to find the optimal number of RF chains that maximizes the system sum-rate for random antenna selection. Let $S$ denotes the cardinal of the set of selected antennas

\begin{equation}
 \label{eq:10}
   S=\sum_{n=1}^{N}\alpha_n
\end{equation}

The sum-rate using conjugate beamformer precoding is asymptotically approximated in~\cite{over2,mrt}. When $K,S \longrightarrow \infty$, the multi user interference term is approximated by its expectation as

\begin{equation}
 \label{eq:11}
   \frac{1}{K} \sum_{i=1,i\neq k}^{K}\mid\mathbf{h_k}\mathbf{h_i}^H\mid^2 \approx \mathbf{E}\{\mid\mathbf{h_k}\mathbf{h_i}^H\mid^2\}=S
\end{equation}

Also $\mathbf{E}\{\mid\mathbf{h_k}\mathbf{h_k}^H\mid^2\}=S^2$ and the term $\eta$ can be approximated as $\eta=\parallel\bm{H}^{H}\parallel_{F}^2\approx K.S$ as in~\cite{over2}. The approximations are validated by simulations in Section V. So, the average $\overline{SINR_k}$ for any user $k$ (where $k = 1, ..., K$) over the channel realizations is approximated as

\begin{equation}
 \label{eq:12}
\overline{{SINR}_k} \rightarrow \frac{\frac{p_{out}}{K}S}{p_{out}+K}
\end{equation}

It is to be noted that the average SINR is the same for all the users. By replacing the circuit power consumption constraint, we obtain the sum-rate averaged over the channel realizations as a function of the total power $p_{max}$, the fixed consumed power $p_c$, the number of users $K$ and the number of selected antennas $S$

\begin{equation}
 \label{eq:13}
   \overline{R}=K.\log_2\left(1+\frac{S(p_{max}-S.p_c)}{K(p_{max}-S.p_c+K)}\right)
\end{equation}

The average sum-rate is a concave function since the second order derivative is negative $\frac{d^2 \overline{{SINR}_k}}{dS^2} < 0$. The optimal number of RF chains that maximizes the average sum-rate is given by

\begin{equation}
 \label{eq:14}
   S^{*}=\text{argmax}_S (\overline{R}),~0 < S^{*} <  p_{max}/p_c
\end{equation}

where $\lfloor p_{max}/p_c\rfloor$ represents the maximum number of RF chains that can be supported by the system due to circuit power constraint. Then, the optimal number of RF chains is derived as

\begin{equation}
 \label{eq:15}
   S^{*}=
   \begin{cases}
\lfloor x\rfloor & \text{if~} \overline{R}(\lfloor x\rfloor) > \overline{R}(\lceil x\rceil)  \text{~or~} \lfloor x\rfloor=\lfloor p_{max}/p_c\rfloor \\
\lceil x\rceil & \text{otherwise}
\end{cases}
\end{equation}
where
\begin{equation}
 \label{eq:30}
  x=\frac{p_{max}+K-\sqrt{K.(p_{max}+K)}}{p_c} < p_{max}/p_c
\end{equation}

The optimal number of RF chains allows to determine the amount of power consumed at the RF chains. Hence, the output transmitted power can be derived and the transmitted power for each user is expressed as

\begin{equation}
 \label{eq:16}
   p_k=\frac{p_{max}-S^{*}.p_c}{K}
\end{equation}

\section{Joint Optimal Number of RF Chains and Power Allocation}

In this section, we derive jointly the optimal number of RF chains and the power allocation among users that maximizes the system sum-rate. An iterative algorithm is proposed, at each iteration for a fixed number of randomly selected antennas, the proposed iterative algorithm computes the optimal power allocation among users. So, the proposed algorithm ends when the sum-rate starts decreasing. Then, the optimal number of RF chains is derived with corresponding optimal power allocation.

Since the power is not allocated equally to users, the approximation in (\ref{eq:11}) cannot be any more used. In this case, we make use of the approximation of the multi user interference term by its expectation. When $K,S \longrightarrow \infty$, the interference term is approximated as in~\cite{mrt2}

\begin{equation}
 \label{eq:17}
 \begin{split}
   \sum_{i=1,i\neq k}^{K}\frac{p_i}{\eta}\mid\mathbf{h_k}\mathbf{h_i}^H\mid^2 \rightarrow \mathbf{E}\{\sum_{i=1,i\neq k}^{K}\frac{p_i}{\eta}\mid\mathbf{h_k}\mathbf{h_i}^H\mid^2\} \\
   =\sum_{i=1,i\neq k}^{K}\ \frac{p_i}{K}
 \end{split}
\end{equation}

This approximation is validated by simulations in Section V. The $SINR_k$ for user $k$ can be expressed as

\begin{equation}
 \label{eq:18}
  {SINR}_k=\frac{p_k \beta_k^2}{S (p_{max}-S.p_c-p_k+K)}
\end{equation}
where $\beta_k^2=\mid\mathbf{h_k}\mathbf{h_k}^H\mid^2$.

The original problem in (\ref{eq:9}) becomes concave since the second order derivative is negative

\begin{equation}
 \label{eq:19}
  \frac{d^2{SINR}_k}{dp_k^2}=-\frac{2 \beta_k^2 (p_{max}-S.P_c+K)}{S (p_{max}-S.p_c-p_k+K)^3} < 0
\end{equation}

Here, the aim is to find jointly the optimal number of RF chains and power allocation among users. For a given $S$, the problem given by (\ref{eq:9}) becomes

\begin{equation}
\label{eq:20}
\begin{aligned}
& \underset{\mathbf{p}}{\text{maximize}}
& & R= \sum_{k=1}^{K} \log_2\left(1+ \frac{p_k \beta_k^2}{S (p_{max}-S.p_c-p_k+K)}\right) \\
& \text{subject to}
& & \sum_{k=1}^{K}p_k+ S.p_c \leq p_{max}.
\end{aligned}
\end{equation}

The associated Lagrangian optimization problem is expressed as

\begin{equation}
\label{eq:21}
\begin{split}
  L(\{p_k\},\mu)= \sum_{k=1}^{K} \log_2\left(1+ \frac{p_k \beta_k^2}{S (p_{max}-S.p_c-p_k+K)}\right) \\
  + \mu \left(\sum_{k=1}^{K}p_k +S.p_c-p_{max}\right)
\end{split}
\end{equation}

By taking the derivative of the above with respect to $p_k$, we can obtain the Karush-Kuhn-Tucker (KKT) system of the optimization problem

\begin{equation}
\label{eq:22}
\begin{split}
p_k^2 (S-\beta_k^2)+ p_k a (\beta_k^2-2.S)\\
+S a^2 +\frac{a \beta_k^2}{\ln(2)\mu}=0
\end{split}
\end{equation}
where $a=p_{max}-S.p_c+K$.

The proposed algorithm solves Eq.(\ref{eq:22}) at each iteration for a fixed number of RF chains $S$  and computes the corresponding optimal power allocation vector among users. The convergence is obtained when the sum-rate starts decreasing ($R_S < R_{S-1}$). The convergence point exists since the sum-rate is a concave function, the second order derivative is negative $\frac{d^2{SINR}_k}{dS^2} < 0$. Hence, the proposed algorithm enables to derive the optimal number of RF chains, the optimal power distribution among users and the maximum achieved sum-rate.

\begin{algorithm}[h]
\caption{Optimal number of RF chains and power allocation algorithm}
\begin{algorithmic}[1]
 \For{$S=1:\lfloor p_{max}/p_c\rfloor$}
\State $\beta_k^2 \gets \mid\mathbf{h_k}\mathbf{h_k}^H(S)\mid^2, k=1:K$
\State $a \gets p_{max}-S.p_c+K$
\State $\Delta_k \gets  a^2(\beta_k^2-2S)^2-4.(S-\beta_k^2).(Sa^2+\frac{\beta_k^2a}{\mu\ln(2)}), k=1:K$
\State  $\sum_{k=1}^{K}\left(\frac{a(2S-\beta_k^2)+\sqrt{\Delta_k}}{2(S-\beta_k^2)}\right)^{+}=p_{max}-S.p_c$ \Comment{find $\mu$ by bisection method}
\State $p_k \gets \left(\frac{a(2S-\beta_k^2)+\sqrt{\Delta_k}}{2(S-\beta_k^2)}\right)^{+}, k=1:K$, power allocation
\State Break If $R_S < R_{S-1}$
\EndFor
\end{algorithmic}
\end{algorithm}

\section{Numerical Results}

In this section, monte carlo simulations are done to validate first the analytical results under equal power allocation assumption. Then, we present numerical results for the proposed algorithm that jointly computes the optimal number of RF chains and power allocation.

We consider that the BS is equipped with 256 antennas serving 10 users. The consumed power at each RF chain is set to 0.05.

\begin{figure}[h]
 \centerline{\includegraphics [width=9cm]  {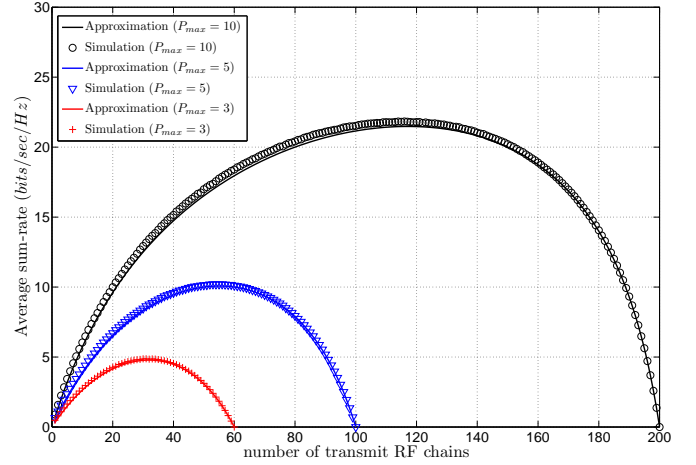}}
\caption{Average sum-rate  under equal power allocation.}
\label{fig1}
\end{figure}

Fig.~\ref{fig1} shows the sum-rate averaged over 1000 channel realizations in function of number of transmit RF chains with conjugate beamformer precoding and under equal power allocation for different available power at the base station. Simulation results confirm the analytical approximation of the sum-rate with conjugate beamformer precoding. Also, we can observe the optimal number of RF chains that maximizes the average sum-rate under equal power allocation for different values of the maximal available power at the base station $p_{max}$.

\begin{figure}[h]
 \centerline{\includegraphics [width=9cm]  {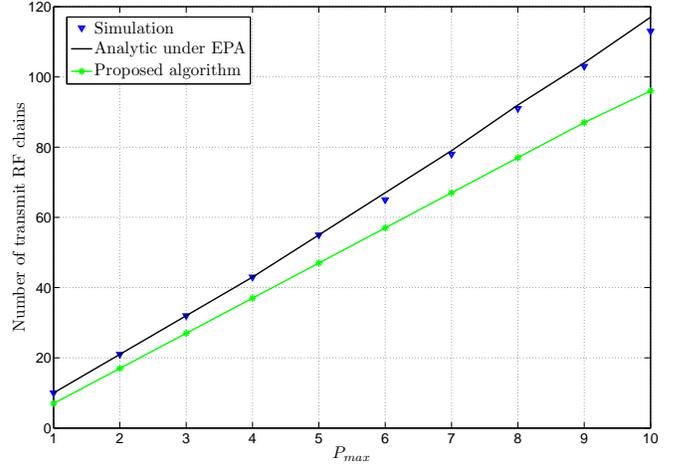}}
\caption{Optimal number of transmit RF chains.}
\label{fig2}
\end{figure}

In Fig.~\ref{fig2} we see the optimal number of RF chains in function of the maximal available power at the base station $p_{max}$ under two scenarios: equal power allocation and the proposed algorithm. First, we confirm that analytical expression of the optimal number of RF chains under equal power allocation fits with simulation results. Next, we can observe the optimal number of RF chains computed by the proposed algorithm when making optimal power allocation. As we see in Fig.~\ref{fig2} we need less number of RF chains when distributing the power optimally among users. Hence, the power consumed at the RF chains significantly decreases.  So, the proposed algorithm offers more available transmitted power for the base station which increases the sum-rate and improves system performance. Also, we can observe that the optimal number of RF chains under the two scenarios linearly increases.

\begin{figure}[h]
 \centerline{\includegraphics [width=9cm]  {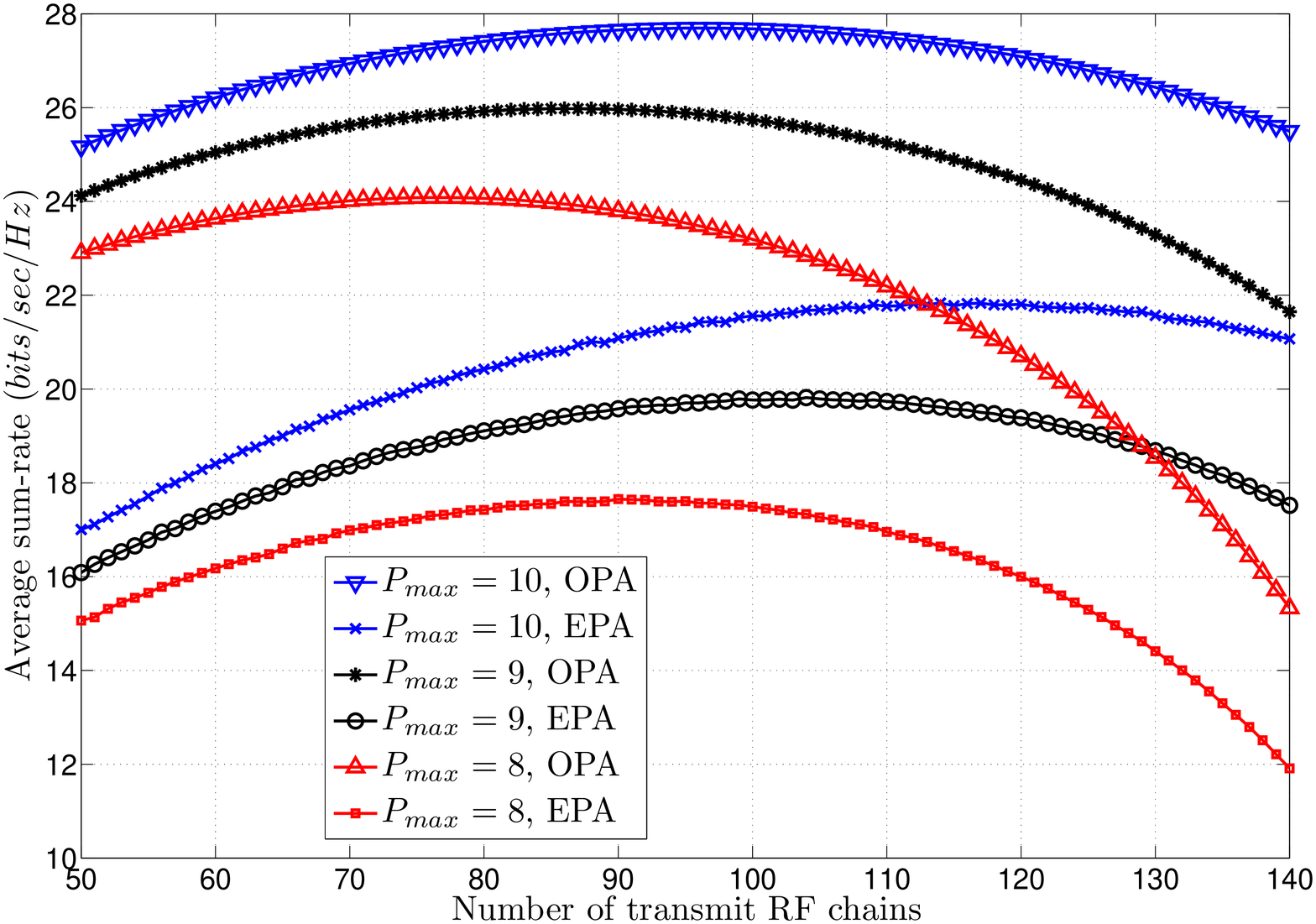}}
\caption{Average sum-rate using the proposed algorithm.}
\label{fig3}
\end{figure}

Fig.~\ref{fig3} shows the obtained average sum-rate under the proposed algorithm as a function of the number of RF chains. It is clear that the achieved sum-rate under the proposed algorithm is significantly higher than the sum-rate obtained under equal power allocation. We have seen the optimal number of RF chains needed for transmission is less when distributing the available power optimally among users. From the circuit power constraint, we deduce that there is more available transmitted power from the base station which increases significantly the achieved sum-rate.

\begin{figure}[h]
 \centerline{\includegraphics [width=9cm]  {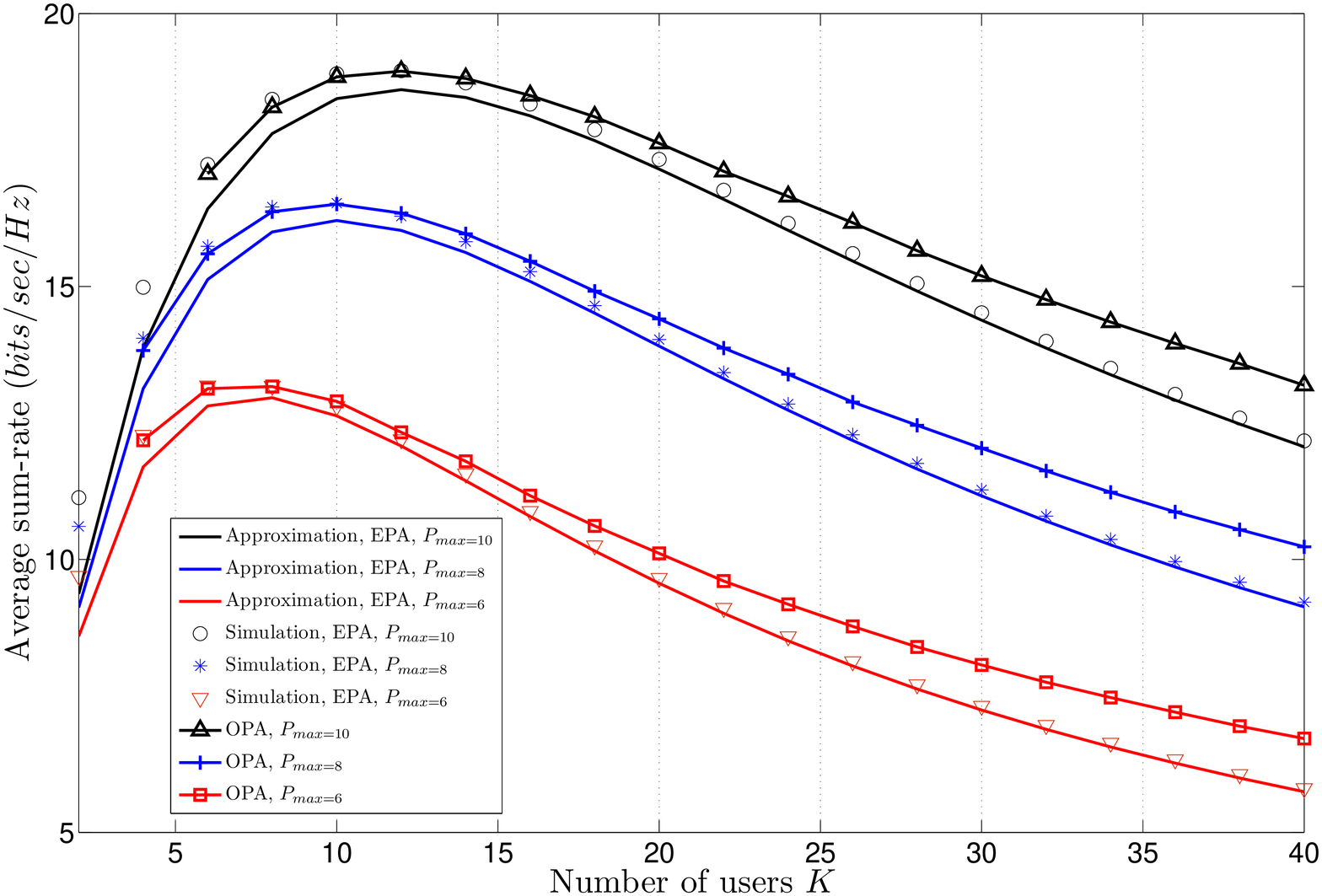}}
\caption{Impact of the number of users $K$ on the system performance (64 RF chains).}
\label{fig4}
\end{figure}

Finally, the impact of the number of users $K$ on the system performance is investigated under equal power allocation and optimal power allocation in Fig.~\ref{fig4}. The analytical approximations of the average sum-rate under equal power allocation fits with the simulation results as $K$ is getting large. The sum-rate decreases because of the multi user interference that exists in conjugate beamforming. Also, the optimal number of users $K$ that maximizes the sum-rate can be observed. It increases when the available power at the BS $p_{max}$ increases, there is more available power to serve a larger number of users. The optimal number of users $K$ that maximizes the sum-rate, is the same for both equal power allocation and optimal power allocation, but the sum-rate is higher when distributing the power optimally among users.

\section{Conclusion}

This paper takes into consideration the circuit power consumption when investigating the downlink transmission in massive MIMO multiuser systems. It confirms first the hypothesis that the maximal achieved sum-rate is not obtained when using all the available RF chains. Hence, under equal power allocation we derive analytical expression of the optimal number of RF chains that maximizes the average sum-rate over channel realizations. Simulation results validate the analytical approximation of sum-rate and the expression of optimal number of RF chains. Then, we propose an algorithm that jointly computes the optimal number of RF chains and distributes the available transmit power optimally among users. Simulations show again that this algorithm allows the base station to use less number of RF chains, so it provides more available transmitted power and achieves high sum-rate.

This paper considers the conjugate beamforming precoding while other beamforming schemes could be also considered. For instance, ZF and regularized ZF will be considered in future work to investigate the impact of beamforming scheme on the optimal number of RF chains. Also, this paper assumes a random antenna selection, whereas a more adequate antenna selection would greatly improve the performances. Antenna selection will certainly impact also the optimal number of RF chains to be activated. Optimal antenna selection and heuristics will be proposed in future works. A joint optimization of power allocation, antenna selection under circuit power consumption model will also be investigated.


\begin{thebibliography}{i}

\bibitem{five} F. Boccardi, R. W. Heath Jr., A. Lozano, T. L. Marzetta and P. Popovski, ``Five Disruptive Technology Directions for 5G,'' \textit{IEEE Commun. Mag.}, vol. 52, no. 2, pp. 74-80, Feb. 2014.

\bibitem{five2} E. G. Larsson, O. Edfors, F. Tufvesson and T. L. Marzetta, ``Massive MIMO for Next Generation Wireless Systems,'' \textit{IEEE Commun. Mag.}, vol. 52, no. 2, pp. 186-195, Feb. 2014.

\bibitem{over1} L. Lu, G. Ye Li, A. Lee Swindlehurst, A. Ashikhmin and R. Zhang, ``An Overview of Massive MIMO: Benefits and Challenges,'' \textit{IEEE J. of Sel. Topics in Signal Processing}, vol. 8, no. 5, Oct. 2014.

\bibitem{over2} F. Rusek, D. Persson, B. K. Lau, E. G. Larsson, T. L. Marzetta, O. Edfors and F. Tufvesson, ``Scaling up MIMO: Opportunities and Challenges with Very Large Arrays,'' \textit{IEEE Signal Processing Mag.}, vol. 30,  no. 1, Jan. 2013.

\bibitem{res1}  H. Q. Ngo, E, G. Larsson and T. L. Marzetta, ``Energy and Spectral Efficiency of Very Large Multiuser MIMO Systems,'' \textit{IEEE Trans. on Commun.}, vol. 61, no. 4, Apr.2013.

\bibitem{res2} A. L. Anderson and M. A. Jensen, ``Beamforming in Large-Scale MIMO Multiuser Links under a Per-Node Power Constraint,'' \textit{IEEE Int. Symp. on Wireless Commun. Systems (ISWCS)}, pp. 821-825, Aug. 2012.

\bibitem{res3}  A. L. Anderson and M. A. Jensen, ``A Generalized Sum-Rate Optimizer for Cooperative Multiuser Massive MIMO Link Topologies,'' \textit{IEEE Acces}, 0.1109/ACCESS.2014.2347241, vol. 2, September 2014.

\bibitem{as1}  A. Liu and V. K. N. Lau, ``Joint Power and Antenna Selection Optimization in Large Cloud Radio Access Networks,'' \textit{IEEE Trans. on Signal Processing}, vol. 62, no. 5, March 2014.

\bibitem{as2}  X. Guozhen, L. An, J. Wei, X. Haige and L. Wu, ``Joint User Scheduling and Antenna Selection in Distributed Massive MIMO Systems with Limited Backhaul Capacity,'' \textit{China Communications}, vol. 11, Issue: 5, May 2014.

\bibitem{as3}  M. Benmimoune, E. Driouch, W. Ajib and D. Massicotte, ``Joint Transmit Antenna Selection and User
    Scheduling for Massive MIMO Systems,'' \textit{ IEEE Wireless Commun. and Networking Conf. (WCNC)},  March 2015.

\bibitem{cir2}  H. Li, L. Song, D. Zhu and M. Lei, ``Energy Efficiency of Large Scale MIMO Systems with Transmit Antenna Selection,'' \textit{IEEE Int. Conf. on Commun. (ICC)}, pp. 4641-4645, June 2013.

\bibitem{cir1}  Y. Pei, T.-H. Pham and Y.-. Liang, ``How Many RF Chains are Optimal for Large-Scale MIMO  Systems When Circuit Power is Considered?,'' \textit{IEEE Global Commun. Conf. (GLOBECOM)}, pp. 3868-3873, Dec. 2012.

\bibitem{mrt}  H. Yang and T. L. Marzetta, ``Performance of Conjugate and Zero-Forcing Beamforming in Large-Scale Antenna Systems,'' \textit{IEEE J. of Sel. Areas in Commun.}, vol. 31, no. 2, Feb. 2013.

\bibitem{wf}  E. Driouch and W. Ajib, ``Efficient Scheduling Algorithms for Multiantenna CDMA Systems,'' \textit{IEEE Trans. on Vehicular Technology}, vol. 61, no. 2, Feb. 2012.

\bibitem{mrt2}  L. Zhao, Hu. Zhao, F. Hu, K. Zheng and J. Zhang, ``Energy Efficient Power Allocation Algorithm for Downlink Massive MIMO with MRT Precoding,'' \textit{IEEE Vehicular Technology Conf. (VTC)}, pp. 1-5, Sept. 2013.

\end{thebibliography}
\end{document}